\documentclass[preprint]{vgtc}               

\graphicspath{{figures/}{pictures/}{images/}{./}} 

\usepackage{times}                     

\usepackage{tabu}                      
\usepackage{booktabs}                  

\usepackage{mathptmx}                  

\onlineid{0}

\vgtccategory{Research}

\vgtcinsertpkg

\title{\toolname: AI-Enabled Climate Change Communication through Personalized and Localized Narrative Visualizations}

\author{Mashrur Rashik\thanks{e-mail: mrashik@cs.umass.edu}\\ %
        \scriptsize University of Massachusetts Amherst %
\and Jean-Daniel Fekete\thanks{e-mail: jean-daniel.fekete@inria.fr}\\ %
     \scriptsize Universit\'e Paris-Saclay, CNRS, Inria %
\and Narges Mahyar\thanks{e-mail: nmahyar@cs.umass.edu}\\ %
     \scriptsize University of Massachusetts Amherst}

\teaser{
  \centering
  \includegraphics[width=\linewidth]{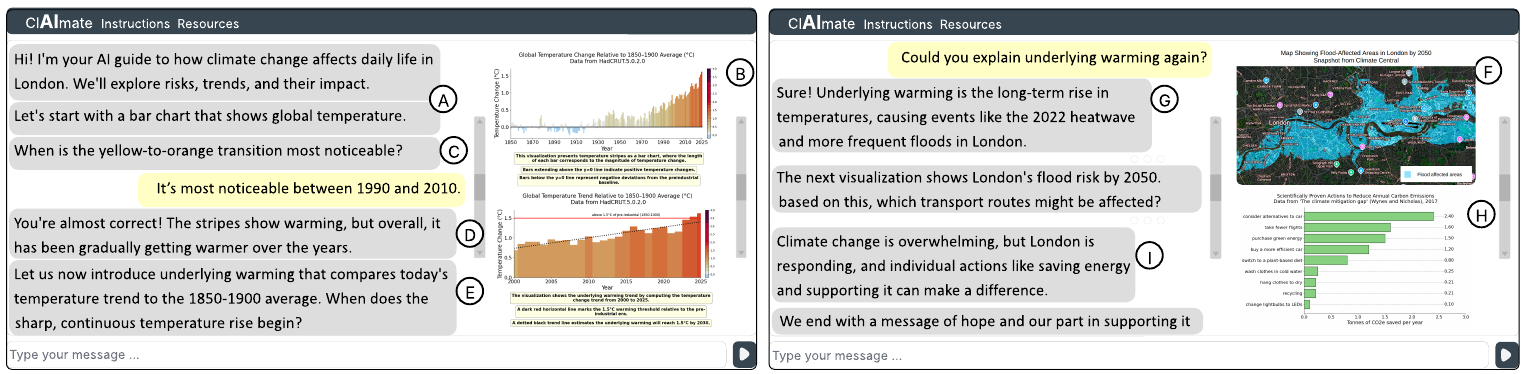}
  \caption{An example walkthrough of the system. 
  (A) \toolname{} begins with a brief introduction and presents the concept of temperature change.
  (B) It presents a visualization of temperature change and uses storytelling to progressively unfold and translate its abstract implications by illustrating their meaning and potential impact on the user’s local area.
  (C) Asks a question to engage the users and reinforce learning.
  (D) An example of a localized response for a London user, with a narrative personalized to their limited climate background. 
  (E) It introduces key concepts such as the difference between temperature change and global warming, and their links to extreme weather. 
  (F) Shows a visualization of flood projections, highlighting one of the key local impacts in the London area.
  (G) \toolname{} answers questions using climate science data and explains ``underlying warming.''
  (H) The final visualization illustrates different human actions and their contribution to carbon emissions. 
  (I) The conversation concludes with a message of hope reinforcing the importance of action plans and their potential to shape a more resilient future.
}

  \label{fig:teaser}
}

\abstract{
\rr{
Communicating climate change remains challenging, as climate reports, though rich in data and visualizations, often feel too abstract or technical for the public. Although personalization can enhance communication, most tools still lack the narrative and visualization tailoring needed to connect with individual experiences. We present \toolname{}, an AI-enabled prototype that personalizes conversation narratives and localizes visualizations based on users' climate knowledge and geographic location. We evaluated \toolname{} through internal verification of factual correctness, a formative study with experts, and a pilot with UK residents. \toolname{} achieved 66\% SNLI accuracy and 70\% FACTSCORE. Visualization experts appreciated its clarity and personalization, and seven out of ten UK participants reported better understanding and local relevance of climate risks with \toolname{}. We also discuss design challenges in personalization, accuracy, and scalability, and outline future directions for integrating visualizations in personalized conversational interfaces.
}

} 


\begin{document}

\newcommand{\ra}[1]{\renewcommand{\arraystretch}{#1}}
\newcommand{\scstart}[1]{\vspace{.3mm} \noindent{\textsc{#1:}}}

\newcommand{\toolname}{\textsc{cl}\scalebox{1.2}[1.1]{AI}mate}

\newcommand{\pheading}[1]{\vspace{4px}\noindent\textbf{#1}}

\newenvironment{tight_itemize}{\begin{itemize} \itemsep -1pt}{\end{itemize}}
\newenvironment{tight_enumerate}{\begin{enumerate} \itemsep -1pt}{\end{enumerate}}

\newcommand{\meanRelevance}[0]{$0.86\pm 0.09$}

\newcommand{\rr}[1]{#1}
%


\maketitle

\section{Introduction}
Climate change poses a profound global threat, yet effectively communicating climate data to the broader public remains difficult.
Despite the extensive efforts and detailed reports produced by organizations such as the Intergovernmental Panel on Climate Change (IPCC), much of this information remains too complex or abstract for lay audiences to engage with meaningfully~\cite{pidcock2021evaluating}. 
Research shows that many people perceive climate change as a distant issue, believing its risks affect others, such as people in other countries or future generations~\cite{spence2012psychological}.
This psychological distance diminishes the sense of urgency and personal relevance needed to motivate action. Therefore, effectively communicating climate change data to broader audiences remains a significant challenge~\cite{budescu2012effective}.

Building on Constructivist Learning Theory~\cite{piaget1950psychology} and Cognitive Load Theory~\cite{sweller1988cognitive}, research on climate change perception~\cite{schuster2024being} and the benefits of storytelling and personalization, 
we posit that personalized storytelling offers a promising approach to reducing psychological distance by making climate data more relatable to the general public.
Prior work in data-driven storytelling has demonstrated that weaving data with personalized narrative can enhance understanding and engagement~\cite{riche2018data}. 
Research also emphasizes the importance of incorporating personal data—such as educational background and lived experience—especially when designing for underrepresented populations for improved communication~\cite{peck2019data}.
Despite the clear benefits of data-driven storytelling for enhancing communication and increasing public engagement, tailoring these narratives to different audiences remains prohibitively expensive. 
To the best of our knowledge, no existing system employs AI to scale personalized data storytelling to support broader audiences in understanding complex scientific data within the context of their local environments and individual backgrounds.

Inspired by the growing body of research on conversational agents for engaging users in complex contexts~\cite{vossen2024effect, salminen2024using} and enhancing visualization comprehension~\cite{gao2015datatone, setlur2016eviza} especially their ability to personalize the conversation~\cite{rashik2025ai}, we introduce \toolname{} (\autoref{fig:teaser}), an AI-enabled climate communication prototype that engages the public by contextualizing complex climate data to their personal and local experiences. We define contextualization as combining personalized narratives and localized visualizations to make climate data more relevant and relatable.
\toolname{} localizes content by presenting visualizations specific to an individual's geographic location and personalizes narrative delivery based on users' education levels and prior climate knowledge obtained through a pre-study questionnaire. It leverages GPT-4, a large language model (LLM), to generate personalized narrative explanations. 
supported by retrieval augmentation that grounds the conversation in credible climate science data.
\toolname{} is designed for the public seeking to better understand climate change impacts and to make scientific data more relatable through personalized explanations. Our focus is on enhancing learning outcomes rather than persuading skeptics or promoting behavior change.
We conducted a pilot study with 10 UK participants 
who found \toolname{} informative, clear, and responsive. 


Our contributions include: 
(1) the design and implementation of \toolname{} \rr{and 
(2)} Insights on key design challenges related to personalization, explanation fidelity, and scalability with suggestions for future work.
\rr{In addition, we provide insights from preliminary evaluation} including: (a) internal system verification to ensure accuracy and context awareness, and (b) a formative usability study with visualization experts and a pilot test with UK residents demonstrating its effectiveness in bringing context and clarity to climate communication. 
\rr{We believe our work is an early attempt at enriching conversational user interfaces with visualizations for more effective communication of complex scientific data to the public.}



\section{Related Work}

This work builds on a growing body of literature at the intersection of data-driven storytelling, conversational agents for data communication, and personalization for engagement and learning outcomes.

\textbf{Data-Driven Storytelling.} Data-Driven Storytelling is the integration of data, narrative structures, and visual representations. It has emerged as a powerful means of communicating complex information in accessible and engaging ways~\cite{riche2018data,shao2024data,segel2010narrative}. In climate change, narratives can make abstract scientific data more relatable and emotionally resonant~\cite{moser2010communicating, mehta2024communication}, bridging the gap between technical content and individual lived experiences~\cite{o2012role}.  
Despite their potential, most climate communication tools rarely implement or systematically evaluate storytelling strategies.




\rr{\textbf{AI for Mitigating Climate Change.} 
AI-enabled tools play a powerful role in mitigating climate change by visualizing climate impacts~\cite{luccioni2021using}, monitoring carbon emissions~\cite{chen2023artificial}, and supporting decision-making for urban planning~\cite{jain2023ai}. 
To ensure meaningful climate action, these AI applications must also align with local contexts and community priorities~\cite{luccioni2021using}. 
Although research explores how AI can leverage climate data for mitigation, existing tools often lack meaningful local context and fail to support bi-directional interaction, limiting users’ ability to engage with and understand the data.
}

\textbf{Conversational Agents for Communicating Complex Data.} Prior work in visualization and HCI has laid important groundwork for conversational data storytelling, including natural language interfaces~\cite{gao2015datatone, setlur2016eviza}, 
and embedded narratives~\cite{shi2023breaking}. 
Most relevant to our work are ChatClimate ~\cite{vaghefi2023chatclimate} and Mena et al.~\cite{mena2025augmenting}, which explores climate question answering through conversational agents.
While these systems demonstrate how conversation can enhance data engagement, they do not integrate personalized interaction.


\textbf{Personalization in Large Language Models.} Recent research also highlights the growing use of large language models (LLMs) for personalization~\cite{kang2023llms, rashik2025ai}. 
Approaches such as vanilla personalization, profile-augmented prompting~\cite{liu2021p}, and retrieval-augmented generation~\cite{samarinas2025beyond} offer scalable ways to tailor outputs to user data. These techniques have been applied across domains, including education~\cite{li2023teach}, healthcare~\cite{rashik2025ai}, and customer service~\cite{pandya2023automating}. 
However, their application in generating personalized narratives for complex climate change data remains unexplored.





\section{\texorpdfstring{\toolname{}}{clAImate}: Design and Implementation}
\rr{\toolname{} is a conversational system designed to deliver personalized text explanations and visualizations about climate change. }
This section introduces \toolname{}, highlights key design considerations, and describes its architecture and implementation.

\subsection{Theoretical Underpinnings}
The overall approach of \toolname{} is guided by two established learning theories: Cognitive Load Theory and Constructivist Learning Theory, which together support the use of personalization to enhance comprehension, engagement, and learning outcomes. Though often treated as distinct, these theories are seen as complementary in visualization research and the design of HCI education tools. Constructivist learning principles have strongly influenced visualization research, particularly in narrative and educational visualizations by grounding data stories in real-world experiences~\cite{segel2010narrative} and connecting them to users' values and prior knowledge~\cite{lee2023affective}. Cognitive load theory has guided visual simplification, annotation, and multimodal delivery~\cite{franconeri2021science}.

Constructivist Learning Theory suggests that learners build new knowledge most effectively when it connects to their existing experiences and contexts~\cite{piaget1950psychology}. By grounding climate narratives in local environments (e.g., temperature changes in one's city or recent heatwaves), \toolname{} fosters deeper understanding and retention by making abstract climate trends more tangible and relatable. Cognitive Load Theory further emphasizes minimizing extraneous cognitive effort to support learning~\cite{sweller1988cognitive}. \toolname{} adapts narrative complexity based on users’ prior knowledge and simplifies explanations for those with less scientific background to reduce cognitive load and help users focus on essential climate concepts.

\subsection{System Overview and Design Considerations}
\rr{
Our system, \toolname{}, was designed with four key design considerations in mind:}

\noindent\rr{\textbf{DC-1: Employ storytelling.} The system should use storytelling to communicate complex climate topics, starting with a high-level concept and gradually introducing key components~\cite{segel2010narrative}.}

\noindent\rr{\textbf{DC-2: Localize visualization and narrative.} The system should localize text and visualization using data specific to the user’s location, making narratives relatable and meaningful~\cite{moser2010communicating}.}

\noindent\rr{\textbf{DC-3: Personalize narrative delivery.} The system should personalize both the depth and complexity of the narrative based on the user's reported education level and climate change knowledge, ensuring the content is comprehensible and engaging~\cite{vaghefi2023chatclimate}.}

\noindent\rr{\textbf{DC-4: Maintain focused narrative.} 
The system should maintain focus on the narrative, providing factual answers based on verified climate data, and then steering the conversation back~\cite{rashik2025ai}.}
\rr{The following outlines how the system supports each:}

\paragraph{DC-1: Employ Storytelling.}
\rr{\toolname{} adopts an interactive slideshow storytelling approach~\cite{segel2010narrative}, where each step in the dialogue presents a visualization, descriptive text, and a related question. The system's conversational feature allows users to further ask questions related to the visualization and its topic. \toolname{} follows a progressive narrative unfolding structure, which is inspired by IPCC reports that guide users through a structured sequence (e.g., past → present → projected futures → actionable steps). \toolname{} begins by explaining its objective to help users explore climate risks and outcomes through a step-by-step narrative grounded in scientific evidence. It then presents observed warming trends, builds on this by explaining their impacts, exploring potential future scenarios, and concluding with clear, actionable steps individuals can take to combat climate change.}

\rr{The narrative includes nine visualizations (see supplementary materials), including a stripe chart displaying temperature observations from 1850 to 2025. We selected stripe charts because climate organizations, including the IPCC, widely use them to communicate temperature change. Despite their wide use to illustrate warming trends, these charts can be difficult for non-experts to interpret since they rely solely on color to show magnitude and do not include important markers, such as the critical 1.5°C threshold. To address these limitations, we created three versions of the stripe chart. The three charts present the data as a bar chart, where the lengths of the bars reflect the magnitude of the temperature change. The first chart covers the temperature change from 1850 to 2025, while the next two zoom in on the period from 2000 to 2025. These charts highlight recent trends, mark critical temperature thresholds, and display extreme events.
To reduce cognitive load~\cite{sweller1988cognitive}, we moved some of the chart descriptions to annotations on the charts.
}

\rr{\toolname{} then uses a stacked bar chart to show the number of natural disasters over time, providing evidence of climate change through events that users are familiar with and can easily relate to. Stacked bar charts make it easy to see both overall trends and how each type contributes over time~\cite{wilke2019fundamentals}. Next, \toolname{} highlights floods, the most common natural disaster risk in the UK, and presents a map showing flood risk in the user’s city. Local maps effectively highlight risk across different regions and are recognized as easy to relate to~\cite{peck2019data}. After presenting local flood risks, \toolname{} highlights how rising sea levels contribute to flood risks and shows observed and projected sea level rise through 2100 with a line chart. \toolname{} returns to the temperature stripe chart and extends it beyond 2025 to show temperature projections through 2100 under different emission scenarios. Finally, the narrative concludes by presenting various actions individuals can take and the amount of emissions each action can save per year. This provides users with practical steps to help combat climate change and reinforces the possibility of a lower-emission future.}

\paragraph{DC-2: Localize Visualization and Narrative.} 
\rr{To implement localization, \toolname{} collects the user's city and country in the UK before starting the conversation (see supplementary materials) and generates narratives and visualizations tailored to that region. For localization, we used London-specific data in this prototype to tailor charts and descriptions for the London area. However, \toolname{} is designed to scale to other regions by incorporating local datasets.}

\paragraph{DC-3: Personalize Narrative Delivery.} \rr{At each step of the conversation, \toolname{} personalizes the descriptive text of the visualizations based on the user’s educational background and climate knowledge, which are collected beforehand through a pre-study questionnaire. To generate these personalized descriptions, \toolname{} uses an LLM that adapts the narrative to each user. In our implementation, we chose GPT-4 as our LLM, due to its accuracy in preserving information and context.}

\paragraph{DC-4: Maintain Focused Narrative.}
\rr{To keep the narrative focused, \toolname{} continuously tracks the user's intent at each step. If the user responds to questions related to a visualization, \toolname{} advances the narrative. However, if the user asks a question, \toolname{} provides an answer before smoothly returning to the main conversation, ensuring the narrative stays on track.}

\subsection{Data Source Integration}
\rr{We grounded \toolname{}'s responses in trusted climate data from NASA, IPCC, and the Met Office. Focusing on the United Kingdom, we identified temperature change and flooding as key risks and selected relevant datasets to build our visualizations and narratives. Historical temperatures (1850–2025) came from the Met Office's HADCRUT5 dataset, future projections (2025–2100) from the IPCC AR6 Synthesis Report, and flood projections from the Met Office and NASA. To generate scientifically accurate responses, we used the IPCC AR6 reports.}

\subsection{System Implementation}
\rr{
At each step of the conversation, \toolname{} presents a visualization, a descriptive text, and a comprehension question to the user. 
When the user responds to the question, \toolname{} uses in-context learning, an approach proven effective for intent prediction~\cite{rashik2025ai}, to determine whether the user is answering the posed question or asking their own.
If \toolname{} determines that the user has answered the posed question, it moves forward in the narrative (\textit{DC-4}).}

\rr{%
If \toolname{} determines that the user is asking their own question or seeking additional information, it searches a trusted database of climate reports using a retriever~\cite{kuzi2020leveraging}. The retriever is paired with an approximate nearest neighbor (ANN) service for scalability. The retriever identifies the most relevant information, which is then used by an LLM to generate an accurate response (\textit{DC-4}). 
Before delivering this response, \toolname{} verifies its accuracy against the original data from trusted sources using a natural language inference (NLI) model. 
If the response does not meet the required accuracy threshold of 0.5~\cite{samarinas2025beyond}, \toolname{} generates a new one. After responding, \toolname{} returns to the main conversation (\textit{DC-4}). 
Users can access \toolname{} through a Flask web application (see \autoref{fig:teaser}), which stores conversations in a MongoDB database for speed and reliability and uses HTTPS APIs for secure communication with the backend.}


\section{Preliminary Evaluation}
We evaluated ClAImate across three axes: (1) factual accuracy via FACTSCORE (70\%), (2) formative study, and (3) pilot deployment with UK residents. These evaluations validate the system's technical accuracy, usability, and communication effectiveness and helped us to further reiterate and improve \toolname{}.

We began by assessing factual accuracy, focusing on both the verification and the correctness of LLM-generated responses. 
To evaluate our response verification, we used the SciTail and SNLI datasets~\footnote{https://huggingface.co/datasets}, that are commonly used for these tasks. We compared two methods: a NLI model (DeBERTa V3) and cosine similarity using T5-base, both selected based on strong performance in prior work~\cite{samarinas2025beyond}.
On the test sets, DeBERTa V3 achieved 60\% accuracy on SciTail and 66.4\% on SNLI, outperforming cosine similarity (33.7\% and 39.6\%, respectively). While top NLI models with 3B parameters reach 94.7\% accuracy, our 400M-parameter NLI model performed competitively, making it a suitable scalable choice for verification.
We used the ClimateQA~\footnote{https://huggingface.co/datasets/Ekimetrics/climateqa-questions-3k-1.0} dataset of 3,426 climate-related questions to compute FACTSCORES~\cite{min2023factscore}, which measures the factual accuracy of the LLM-generated responses. We calculated scores of the responses that were verified, with an average score of 70\%. In comparison, human judgments typically achieve 88\% on similar tasks.

Although \toolname{} is intended for general audiences, we first conducted a formative usability study with seven visualization researchers to leverage their domain expertise in evaluating narrative clarity, visualization effectiveness, and system functionality. 
\rr{The researchers were not involved in the research and were recruited from a visualization lab through an email announcement. 
They voluntarily participated in the study.
Each participant completed informed consent, read instructions, and filled out a pre-study questionnaire on their location, education, and climate background. After interacting with \toolname{}, they completed a post-study questionnaire (see supplementary materials) about their engagement, satisfaction, and feedback. Finally, we had a one hour focus group meeting with the participants where they shared additional feedback.
}
Participants highlighted both strengths and areas for improvement in the tool's design and interaction flow.
Several participants appreciated the responsiveness and personalization by saying:  
``\textit{It gave me a lot of grounded information I was not familiar with}.''  
``\textit{I liked that the system provided facts and personalized the narrative}.'' and
``\textit{I liked that you are using a system to talk to people and show visualization at the same time}.''
However, others noted limitations in the system's flexibility and dialogue design:  
``\textit{Sometimes it feel like the system doesn't care about my question, just wants to return to the conversation}.''  
``\textit{I liked the fact that the system was asking questions to help me stay engaged; however, it felt like responses were not always tailored to my questions}.''  
``\textit{The interface brings too much text sometimes too fast}.''

    
\rr{To improve responsiveness, we refined the prompt and response generation to acknowledge user input better and provide more natural replies. Furthermore, to reduce textual overload, we segmented visualization descriptions with GPT-4 and added them as annotations directly on the visualizations.}
We then conducted a pilot study with 10 participants ($P\#$) from the UK area, recruited through Prolific.
\rr{The procedure for this study followed our initial formative study, except participants were not interviewed afterwards.}
The recruitment and procedure for the pilot were approved by an Institutional Review Board (IRB).
Our analysis of conversation transcripts and questionnaire responses showed that participants gained a better understanding of climate risks after using \toolname{}.
Most participants (seven out of ten) found the conversation narrative and the visuals relatable and personal to their local experiences.
They also appreciated the informativeness of the responses (P1, P4-6, P10) and clarity (P1-2, P6, P7) 
of \toolname{}.

\rr{We further improved the system based on feedback from pilot participants, and we plan to conduct a summative study with UK residents, with four conditions: (a) a baseline version with neither personalization nor localization, (b) localization only, (c) personalization only, and (d) both localization and personalization. 
Our goal is to evaluate how different components of the system contribute to engagement, recall, and learning compared to a baseline. Our measures for evaluation are informed by interdisciplinary research on engagement, affective learning~\cite{lee-robbins2022affective}, and climate communication~\cite{van2021development}.
}



\section{Challenges and Opportunities}
We discuss challenges and opportunities in \toolname{}'s design and implementation to inform future climate communication systems.

\textbf{Personalization, localization, and design limitations.}
Personalized and localized narratives offer a powerful way to make climate communication more relatable. Yet, designing for diverse user backgrounds, cognitive styles, and sensitivities remains a challenge~\cite{moser2010communicating}. 
While \toolname{} integrates location-aware data and adapts narratives to the user's knowledge, such personalization is inherently constrained. First of all, not all preferences or contexts can be anticipated, and adaptive systems risk reinforcing existing beliefs or introducing bias. Second, LLMs still struggle to generate precise, interpretable visualizations~\cite{wu2024automated, ye2024generative, mena2025augmenting}.
\toolname{} represents an early step toward building AI-enabled systems for personalized climate communication.
Our work raises open questions about balancing narrative control with personalization and the potential risks of using generative models in scientific messaging.
And how can we responsibly scale such systems while respecting users' diverse values, literacies, cognitive capacities, and affective responses?

\textbf{Scalability.}  
\toolname{} indicates the potential of personalized, conversational systems to scale climate communication by leveraging LLMs with RAG and ANN services to process large-scale climate data for different regions efficiently.
Data-driven storytelling, previously resource-intensive, can now be delivered on-demand, reaching diverse users across regions, languages, and levels of climate literacy~\cite{vaghefi2023chatclimate, mehta2024communication}. 
Through the integration of LLMs and RAG, \toolname{} dynamically delivers localized visualizations and narrative explanations tailored to the user's background and location. 
We integrated an ANN service into \toolname{} to accelerate response generation and allow the system to scale effectively by grounding conversations in large collections of climate science data.
This shift from static messaging to adaptive opens new possibilities for inclusive, timely, and personalized climate storytelling~\cite{moser2010communicating}.
\rr{A scalable system, such as \toolname{}, can help raise public awareness, facilitate community discussions about local climate action plans, and supplement media articles or reports by providing personalized context and interpretation of climate impacts based on specific geographic locations.}
Nevertheless, this approach relies heavily on the availability of curated, structured scientific data. In many regions of the world, reliable or localized datasets are scarce or unavailable, limiting the scalability and application of \toolname{}. 
Scaling to other domains—or even to other subfields of climate science—therefore requires new strategies for domain adaptation, content verification, and interaction design that can account for such data disparities.
Other remaining questions include: How can such systems maintain relevance and accuracy while adapting to evolving data and user needs? and What strategies can be implemented to ensure equitable access and safeguard user privacy?


\textbf{Algorithmic Accuracy.} 
By leveraging LLMs, RAG, and ANN search, \toolname{} demonstrates a scalable approach to personalized climate storytelling~\cite{vaghefi2023chatclimate, mehta2024communication, richardson2023integrating}. 
One important consideration for such systems is maintaining factual accuracy: personalized responses generated by LLMs must ensure accuracy and avoid oversimplification, particularly when incorporating sensitive or locally specific data~\cite{samarinas2025beyond, vaghefi2023chatclimate}. To support accuracy and reliability, \toolname{} incorporates a verification that aligns LLM responses with scientific sources and allows adjustments to personalization based on user preferences. While this offers an initial step toward trustworthy interaction, future work should examine ways to surface uncertainty and address issues of trust and fairness, enabling a greater degree of personalization.
Open questions remain, such as: How can we further improve the ability of AI-enabled systems to handle uncertainty and ensure the responsible communication of climate data, especially in diverse  geographical contexts?


\textbf{Limitations.}
\toolname{} relies on pre-rendered visualizations, as LLMs cannot yet generate accurate climate visualizations dynamically~\cite{mena2025augmenting}. We initially used the user's location to select relevant climate datasets and generate visualizations with a Matplotlib-based chart generator. However, the limited availability of rich, location-specific climate data posed a key limitation for automatic visualization generator. 
Furthermore, while \toolname{} answers user questions, the system often steers the conversation back to cover the required educational content
which may limit flexibility.
Personalization in conversational agents is a broad and complex design space. We focused on implementing and evaluating a few personalized aspects related to learning outcomes. 
Future systems should provide greater user agency to personalize content, including narrative style, length, and other preferences.
The system architecture is scalable to other regions, but its performance depends on the granularity and detail of local data.

\section{Conclusion}



\toolname{} offers an instantiation of how AI-driven narrative visualizations can ground complex climate data in users’ local contexts. Integrating natural language understanding, augmented retrieval, and storytelling principles, the prototype delivers context-aware climate narratives. \toolname{}'s factual correctness, functionality, and clarity were evaluated through internal testing, a formative study, and a pilot, all of which yielded promising results and highlighted the system's value in personalizing and localizing climate communication. 
\rr{\toolname{} is a step toward systems that integrate communicative visualization with personalized LLM-driven interaction to inform the design of more accessible and engaging climate communication tools.}

\bibliographystyle{abbrv-doi}

\bibliography{bibliography}
\end{document}